\def\year{2025}\relax
\documentclass[letterpaper]{article} % DO NOT CHANGE THIS
\usepackage{aaai25}  % DO NOT CHANGE THIS
\usepackage{times}  % DO NOT CHANGE THIS
\usepackage{helvet}  % DO NOT CHANGE THIS
\usepackage{courier}  % DO NOT CHANGE THIS
\usepackage[hyphens]{url}  % DO NOT CHANGE THIS
\usepackage{graphicx} % DO NOT CHANGE THIS
\usepackage{natbib}  % DO NOT CHANGE THIS AND DO NOT ADD ANY OPTIONS TO IT
\usepackage{caption} % DO NOT CHANGE THIS AND DO NOT ADD ANY OPTIONS TO IT
\DeclareCaptionStyle{ruled}{labelfont=normalfont,labelsep=colon,strut=off} % DO NOT CHANGE THIS
\usepackage{algorithm}
\usepackage{algorithmic}
\usepackage[toc]{appendix}
\usepackage{longtable}
\usepackage[absolute,overlay]{textpos}
\usepackage{microtype} 

\usepackage{newfloat}
\usepackage{listings}
\floatstyle{ruled}
\newfloat{listing}{tb}{lst}{}
\floatname{listing}{Listing}

\usepackage{enumitem}
\usepackage{tabularx}
\usepackage[table,xcdraw]{xcolor}
\usepackage{amsmath}
\usepackage{graphicx}
\usepackage{tikz}
\usetikzlibrary{backgrounds,fit,positioning}
\usepackage{standalone} 
\usepackage{tikzscale}
\usepackage{enumitem}
\usepackage{subcaption}
\usepackage{booktabs}
\usepackage{cleveref}
\setlist{nosep,after=\vspace{\baselineskip}}
\usepackage{pgfplots}
\usepackage{array}
\usetikzlibrary{patterns}
\usepackage{xxcolor}
\definecolor{yellow}{rgb}{0.85,0.65,0.1}
\usepackage{cuted}
\usepackage{array,caption}
\usepackage[justification=centering]{caption}
\usepackage{quoting}
\usepackage{colortbl}
\usepackage{dblfloatfix}
\usepackage{float}
\usepackage{spverbatim}
\usepackage[skip=0.333\baselineskip]{caption}
\newlength\mylen
\newif\ifnotes
\notestrue

\usepackage{lscape}
\usepackage{subcaption}
\usepackage{multirow}
\usepackage{makecell}

\usetikzlibrary{calc}
\tikzset{caption/.style={execute at end picture={\path
let \p1=($(current bounding box.east)-(current bounding box.west)$) in
(current bounding box.south) node[below, text width=\x1-4pt,align=center] 
{\captionsetup{font=LARGE,labelfont=LARGE}\captionof{figure}{\LARGE #1}};}}}

\newif\ifnotes
\notestrue

\title{A Crowdsourced Study of ChatBot Influence in Value-Driven Decision Making Scenarios}
\author {
    Anthony Wise,\textsuperscript{\rm 1}
    Xinyi Zhou, \textsuperscript{\rm 1}
    Anind Dey, \textsuperscript{\rm 1}
    Martin Reimann \textsuperscript{\rm 2}
    Leilani Battle, \textsuperscript{\rm 1}
}
\affiliations {
    \textsuperscript{\rm 1} University of Washington\\
    \textsuperscript{\rm 2} University of Arizona\\
    anthw85@cs.washington.edu, xzhou@cs.washington.edu,
    anind@uw.edu,
    reimann@arizona.edu,
    leibatt@cs.washington.edu
}

\begin{document}
\maketitle

\begin{abstract}
Similar to social media bots that shape public opinion, healthcare and financial decisions, LLM-based ChatBots like ChatGPT can persuade users to alter their behavior. Unlike prior work that persuades via overt-partisan bias or misinformation, we test whether framing alone suffices. We conducted a crowdsourced study, where 336 participants interacted with a neutral or one of two value-framed ChatBots while deciding to alter US defense spending. In this single policy domain with controlled content, participants exposed to value-framed ChatBots significantly changed their budget choices relative to the neutral control. When the frame misaligned with their values, some participants reinforced their original preference, revealing a potentially replicable backfire effect, originally considered rare in the literature. These findings suggest that value-framing alone lowers the barrier for manipulative uses of LLMs, revealing risks distinct from overt bias or misinformation, and clarifying risks to countering misinformation.
\end{abstract}

\section{Introduction}

The use of ChatBots\footnote{A ChatBot is defined as ``a computer program designed to simulate conversation with human users, especially over the internet...it uses Natural Language Processing (NLP) and sentiment analysis to communicate in human language by text or oral speech with humans or other ChatBots'' \cite{ADAMOPOULOU2020, khanna2015}.} powered by Large Language Models (LLMs) has grown rapidly, with ChatGPT surpassing 100 million users within two months of launch \cite{openai2024gpt4technicalreport}. Beyond basic inquiries, these systems aid in complex decision making for healthcare, finance, and urban planning. For instance, LLMs that analyze large clinical datasets have shown promise in generating treatment recommendations \cite{nah2023}.

As LLM-based ChatBots expand into sensitive domains such as public policy and finance, concerns arise that they can be exploited to spread disinformation or mislead users, akin to social media bots \cite{bada2019cyber, sison2023chatgpt, jakesch2023co, khurana, ADAMOPOULOU2020, kenny2024duped, wischnewski2024agree}. Work on zero-shot persuasive ChatBots shows how AI tools (originally intended for positive applications) can curb conspiracy beliefs \cite{Costello2024DurablyRC}, but can also be repurposed to manipulate user attitudes within synthetic consumer-style domains \cite{furumai2024}. However, the extent and ramifications of these manipulations are still unclear, given that attitude shifts can potentially lead to decision making and behavior changes as a result. In psychology, attempts at persuasion can sometimes \emph{backfire}, a phenomenon that produces an attitude change opposite to the intended direction, though large-scale replications suggest \emph{correction} backfires are rare \cite{Nyhan2010WhenCF, Wood2019TheEB}. Our study differs in that, rather than correcting factual error, we evaluate \emph{value-framed} persuasion on a salient policy choice (US defense spending) \cite{chanDebunk2017}. Further, \emph{we seek to understand whether a person's values (e.g., political stance) modulate their susceptibility to ChatBot manipulations, and the rarity of backfire effects under value-misalignment.}

Towards closing this empirical gap, we present a crowdsourced study where 336 participants interact with two LLM-based ChatBots that attempt to manipulate them according to political ideologies, compared to a neutral control ChatBot.
Our study addresses the following research questions:
\begin{itemize}[leftmargin=0.9em]
    \item \textbf{RQ1}: Do value-driven ChatBots influence users' defense spending preferences?
    \item \textbf{RQ2}: How do participants' values interact with the ChatBot manipulations and is there potential for a value-misaligned backfire effect?
\end{itemize}

To answer these questions, our study design emphasizes a specific yet impactful policy preference scenario, where we ask participants whether to increase or decrease the US government's budget for defense spending. Participants select their budget preference, interact with a pro-Decrease, neutral, or pro-Increase ChatBot, then decide whether to keep or modify their original preference.
Further, we explore whether targeted injection prompts\footnote{\emph{Injection prompts} (a.k.a. prompt injection attacks) are adversarial inputs that ``inject adversarial text or instructions'' to steer an LLM away from its original directions, effectively overriding intended controls; attacks can be \emph{direct} (malicious instructions given as user input) or \emph{indirect} (instructions hidden in external content the model reads, which we focus on for our study), which ``blur the line between data and instructions'' \cite{greshake2023not,Liu2023PromptIA,zhang2024goal}.} may cause the underlying LLMs to manipulate the personal values of users \cite{fab2014, johnson2022ghost}.
Specifically, by prompting the LLMs to emphasize ideologies such as pacifism or economic conservatism, ChatBots may steer users susceptible to these ideologies toward specific outcomes (e.g., reduced vs. increased government spending). These LLM ``bad actors'' do not need to fabricate facts to persuasively nudge the conversation. By keeping the substantive content consistent, with each ChatBot providing accurate information and no misinformation, and altering only the moral framing, our study offers a simple test of value cues within an LLM conversation.

\begin{figure*}[t]
    \centering
    \includegraphics[width=0.45\textwidth]{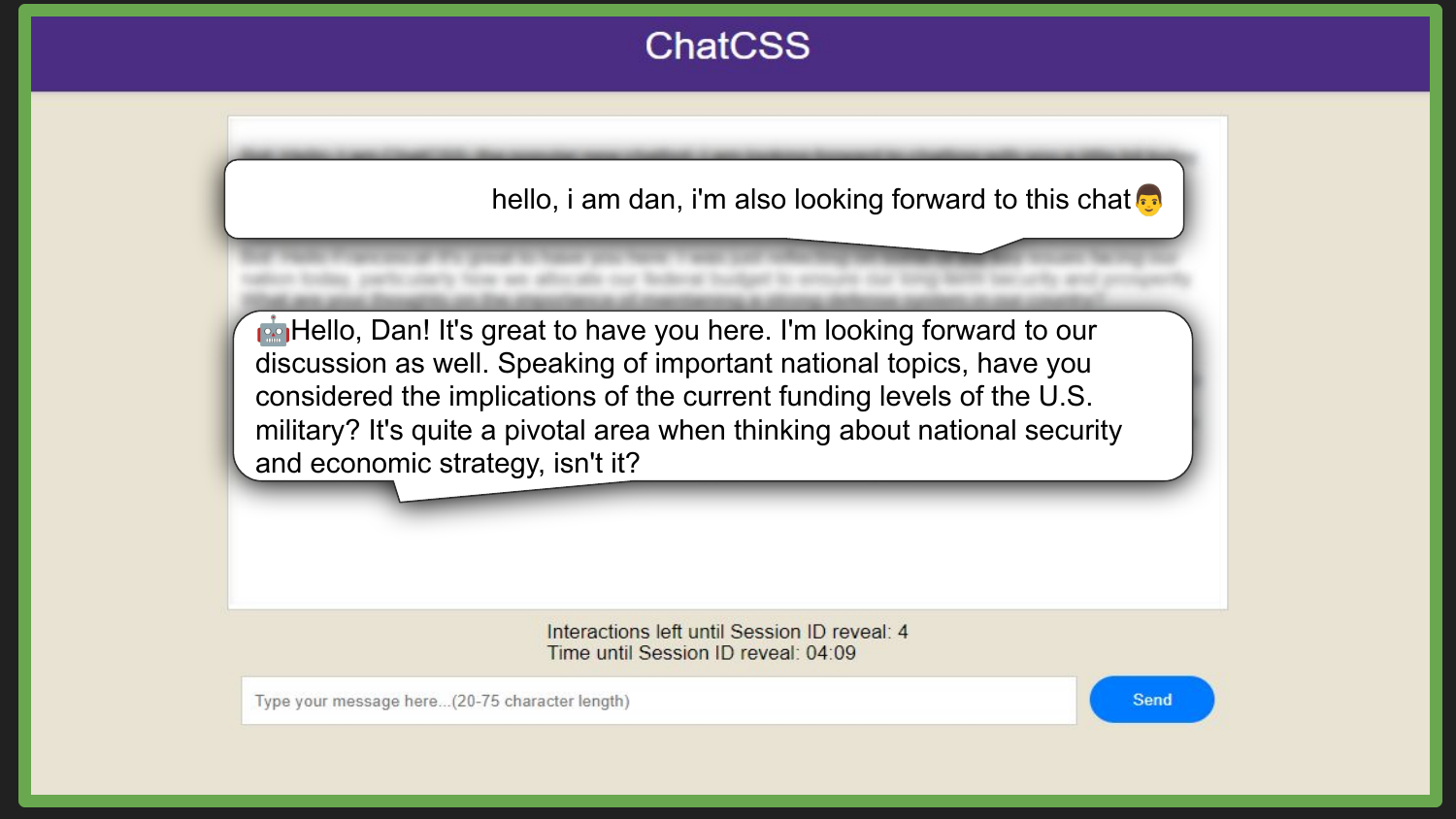}
    \caption{An example of how a compromised ChatBot can subtly manipulate the user to increase spending.
    }
    \label{fig:increase bias}
\end{figure*}

Our study revealed two significant findings. First, when compared to a neutral control ChatBot, 
participants who interacted with the Decrease ChatBot subsequently expressed lower spending preferences, whereas those who spoke with the Increase ChatBot voiced higher ones. A one-way ANOVA confirmed that this divergence is statistically significant and remains significant across participant values (i.e., pacifistic or fiscally conservative). 
Second, in a thematic analysis of participants' ChatBot conversations, we find that participants tended to be swayed by value-aligned ChatBots (e.g., when political liberals were paired with the decrease spending ChatBot) but were repelled by misaligned ChatBots (e.g., when conservatives paired with this same ChatBot). This shows that value \emph{alignment} reliably persuades, while \emph{misalignment} can provoke a \textbf{backfire effect} \cite{Nyhan2010WhenCF, Wood2019TheEB, SwireThompson2019PublicHA}, where participants chose to exaggerate their original preference (e.g., further decrease or further increase spending), where we observed resistance against the value-misaligned ChatBot.
We found that one-third of conservatives resisted the Decrease ChatBot, and over half of liberals resisted the Increase ChatBot, reinforcing rather than relaxing their positions.

We present this exploratory study, focusing on one substantive domain (defense spending), one focal value-frame (\emph{pacifism} or \emph{economic conservatism}), and controlled message content, where each ChatBot delivered the same accurate facts but was guided by a distinct injection prompt to frame the issue through either \emph{pacifist} or \emph{fiscally conservative} values. The pattern we observe, consistent backfire under value-misalignment, suggests that backfire may be less rare in value-framed human LLM dialogue than in factual correction settings, and motivates follow-up tests across other issues and ChatBot designs.

By examining the nuanced ways that personal values interact with AI guidance, our research contributes an exploration of whether participants are more likely to accept or reject AI recommendations when these align with or challenge their core beliefs.

To summarize, this paper makes the following contributions:

\begin{itemize}[leftmargin=0.9em]

    \item We contribute an experiment design that augments an LLM ChatBot's \emph{value-framing} separately from its \emph{persuasiveness}.

    \item In a controlled experiment with 336 participants, we show that a single LLM conversation can causally shift participants' attitudes in a specific, real world context (US defense spending preferences). A one-way ANOVA confirms the statistical significance of this shift. Our findings motivate future experiments to test the robustness of this effect.

    \item Comparing ChatBot condition with participant preferences reveals that value-alignment reliably persuades, while misalignment may reliably produce a \textbf{backfire effect} attitude change opposite to the ChatBot’s intended direction. Our results provide preliminary evidence and identify testable boundary conditions for future replication.
\end{itemize}
\section{Related Work}
\label{sec:related work}

In this section, we situate our study within four strands of prior work: ethical and trustworthy AI; conversational persuasion and political bias in LLMs; corrections and backfire; and AI for decision support, security, and personalization. Prior work often centers on overt partisanship or misinformation; in contrast, we isolate value-framing without manipulating facts constant to test alignment and misalignment effects on a concrete budget choice, extending concerns about AI influence to subtler value-based cues for participant attitude shifts.

\subsection{Ethical \& Trustworthy AI}
Researchers highlight how AI systems like ChatGPT can pose ethical challenges related to deception, user trust, and misalignment with user expectations \cite{schlag2023, harrington2023trust, khurana, lim2011investigating, lim2009and, Cheng2023TalkWC, glikson2020human, sharma2024generative, Wach2023TheDS}. They emphasize the need for transparency and proactive engagement with users—essential for mitigating deceptive behaviors in AI \cite{zhan2023deceptive, sison2023chatgpt}. Emphasis is placed on aligning user expectations with actual AI capabilities \cite{harrington2023trust} and on building equitable, inclusive systems that address biases across diverse user populations \cite{harrington2023trust}. AI’s role in spreading or combating misinformation also reinforces the importance of preserving public information integrity \cite{huang2024fakegpt}. Discrepancies between user expectations and AI outputs highlight the call for greater transparency from providers \cite{greshake2023not}. Additional work explores AI’s potential to enhance information security awareness \cite{gundu2013ignorance} and to influence emotional and decision making processes \cite{fab2014}. Recent work has shown that LLMs can effectively reduce the belief in conspiracy theories through personalized and fact-based dialogues, even among the most devout believers of conspiracy theories \cite{Costello2024DurablyRC}.
Similarly, recommendation algorithms with calibrated political bias can actually reduce ideological polarization in online discussions, showing how subtle value-framing within an AI system can positively influence broader policy attitudes \cite{Peters2022AlgorithmicPB}.
However, well reasoned counterarguments can still inadvertently reinforce misinformation \cite{chanDebunk2017}, and even subtle algorithmic political bias can entrench users within ideological information loops \cite{Franke2022}, highlighting the nuanced design choices required for automated corrective conversations. In another recent experiment, Glickman et al. show that repeated human–AI interaction can create a bias amplifying feedback loop, where algorithms trained on slightly biased human data become biased themselves, and subsequent interaction leaves humans more biased over time; conversely, accurate AI improves human accuracy. Influence is shaped both by the AI’s outputs and by whether it is perceived as ``AI'' versus ``human,'' and participants often underestimate the AI’s impact by raising transparency and framing risks in everyday tools \cite{Glickman2024HowHF}. This mechanism complements our focus on value-framing by showing how seemingly informational assistance can quietly reshape judgments across domains. In our work, we extend ethical considerations by empirically demonstrating how value-framed ChatBots can influence user decisions in targeted ways, revealing underexplored risks of persuasive AI in civic contexts.

\subsection{Conversational AI Persuasion \& Political Bias}
A recent large-scale conversational experiment indicates that LLM persuasiveness depends more on post-training and prompting than on personalization or raw model scale; information-dense prompting yields the strongest effects, some of which persist weeks later, but greater persuasiveness can coincide with lower factual accuracy. Experimental results situate persuasion risks in concrete design choices (such as reward modeling that increases information density) rather than in scale alone \cite{hackenburg2025leverspoliticalpersuasionconversational}.
Another recent complementary interactive study shows that partisan biased LLMs can shift users’ policy opinions and concrete budget allocations toward the model’s bias (even against prior partisanship) with only weak mitigation from prior AI knowledge and little mitigation from merely recognizing bias. Their work further suggests that bias manifests through differential framing (security vs. fairness frames) rather than distinct persuasion ``techniques,'' accentuating the importance of framing control in conversational systems \cite{fisher-etal-2025-biased}. Our work adds to this literature by isolating value- framing (e.g., fiscal conservatism) delivered by seemingly neutral agents and by measuring downstream shifts in US budget preferences; together, our findings emphasize that subtle framing and information-dense dialogue policies can shape civic judgments without explicit misinformation or overt partisanship.

\subsection{Corrections, Backfire, and the Continued Influence of Misinformation}
\label{related-work:backfire-effect}

Early demonstrations of the ``backfire effect'' suggested that factual corrections embedded in realistic news formats can sometimes entrench misperceptions among ideologically predisposed audiences \cite{Nyhan2010WhenCF}. Subsequent large-scale replications, however, find that belief increase backfires are rare; across thousands of participants and dozens of polarized issues, corrections generally move beliefs toward the facts rather than away from them \cite{Wood2019TheEB}. A complementary meta-analysis shows that the influence of misinformation typically persists, even when corrections work: corrections reduce but do not fully erase misinformation’s effects, with efficacy shaped by message coherence, alignment with audience worldviews, repetition, timing, and source credibility \cite{Walter2019AME}. Reviews of online misinformation further highlight how platform dynamics and information-seeking behavior (e.g., confirmation bias) create conditions for both misperception formation and the limited potency of corrections \cite{SwireThompson2019PublicHA}. Our work does not test factual corrections. Instead, we had participants engage in a value-framed conversation with an LLM ChatBot using only factual information. When the ChatBot’s value-frame opposed a participant’s prior stance, some participants reported increased commitment to their original policy position, which is an example of a backfire effect triggered not by misinformation, but by value-misalignment. We interpret this as a value-threat-driven form of backfire, conceptually distinct from classic belief-based backfire, and more closely aligned with motivated reasoning and moral reframing. The mechanism in our case is not a failed correction but resistance to an opposing value-frame embedded in an otherwise informational dialogue.

\subsection{Enhancing Decision Support and Educational Impact Through AI}

Research reveals how AI can streamline decision making and boost collaboration, from improving group dynamics to adapting real-time interactions \cite{chiang2024, nah2023}. In education, generative AI techniques personalize learning, reshaping traditional teaching approaches \cite{baidoo2023education}. Studies on AI-driven visualization and recommendation systems stress the importance of trustworthy designs that meet users’ decision-making needs \cite{zehrung2021vis, bao2022recommendations}. Our work complements this body of work by showing that AI systems (when framed around values such as fiscal conservatism) can influence not only task-specific outcomes but also high-level policy preferences, such as military budget decisions, thereby expanding the scope of decision-support research into value-sensitive civic applications.

\subsection{Securing and Personalizing AI for Enhanced Emotional Interaction and Reliability}

Prompt injection attacks and language-processing vulnerabilities highlight the need for robust security protocols \cite{zhang2024goal, netgpt, bada2019cyber}. Advancements in AI's natural language understanding aim to improve reliability \cite{cox2023comparing}. Researchers also investigate aligning AI with personality traits—such as extraversion—to enhance user engagement and usability \cite{volkel2022user}. Moreover, AI ChatBots can be manipulated to present selectively biased information, risking undue influence on user decisions \cite{bada2019cyber, kowa09, greshake2023not}. Our work contributes to this literature by revealing that even modest value-framing (without overt bias or misinformation) can significantly shift user attitudes, showing how personalization can be harnessed not just for engagement, but for subtle ideological influence.

Personalizing AI to human nuances requires exploring factors like extraversion and professional workflows \cite{volkel2022user, muiklwijk24}. AI applications in mental health show promise in providing support and interventions \cite{xu2024mental}. Research also focuses on subjective or sentiment analysis for empathetic responses \cite{Wilson2008FinegrainedSA}, while AI's potential to surpass human performance in specific tasks raises questions about trust and transparency \cite{motoki2024more}. Finally, studies on AI-driven digital communication stress the need for responsible discourse-shaping technologies \cite{santurkar2023whose}. Our research builds on this by analyzing not only how ChatBots elicit emotional and reflective responses, but how these emotional cues interact with users' value systems to produce decision shifts (or in some cases, a backfire effect) therefore offering a more comprehensive picture of AI's affective influence on human reasoning.
\section{Methods}
\label{sec:methods}

Given consistent negative results for correction-based backfire effects in the literature, we opted to conduct a small-scale, exploratory experiment to ascertain whether value-framing may have a persuasive effect within a ChatBot environment. Our hope is to collect initial data that may inspire future experiments in this area, or in the case of negative findings, reinforce the conclusions of existing work. To this end, we conduct a single, crowdsourced experiment to explore the effects of value-framed influence on attitude-shifting in ChatBot conversations.

\subsection{Study Procedure}

\begin{figure}[h]
\begin{tikzpicture}
    \begin{axis}[
        width=9.0cm,height=3.5cm,ybar,bar width=4pt,xmin=0, xmax=46,ymin=0, ymax=200,xlabel={Number of Participant Interactions},ylabel={Number of Participants},title={Distribution of Participant Interaction Counts},tick label style={font=\tiny},label style={font=\footnotesize},ylabel style={yshift=-1.4em},xtick={0,2,4,6,8,10,12,14,16,18,20,22,24,26,28,30,32,34,36,38,40,42,44,46},xtick pos=left,
        ytick pos=left
    ]
    \addplot[fill=blue, draw=black] coordinates {
        (0,0)
        (1,11)
        (2,67)
        (3,41)
        (4,22)
        (5,15)
        (6,190)
        (7,67)
        (8,22)
        (9,20)
        (10,15)
        (11,10)
        (12,13)
        (13,5)
        (14,11)
        (15,0)
        (16,1)
        (17,1)
        (18,0)
        (19,0)
        (20,0)
        (21,0)
        (22,0)
        (23,0)
        (24,1)
        (25,0)
        (26,1)
        (27,0)
        (28,0)
        (29,0)
        (30,0)
        (31,0)
        (32,0)
        (33,0)
        (34,0)
        (35,0)
        (36,0)
        (37,0)
        (38,0)
        (39,0)
        (40,0)
        (41,0)
        (42,0)
        (43,0)
        (44,0)
        (45,1)
        (46,0)
    };
    \draw[red, thick, dotted] (axis cs:6,0) --(axis cs:6,200);
    \end{axis}
\end{tikzpicture}
\caption{User interaction distributions where the red dotted line shows most participants made 6 interactions}
\label{fig:inter_hist}
\end{figure}
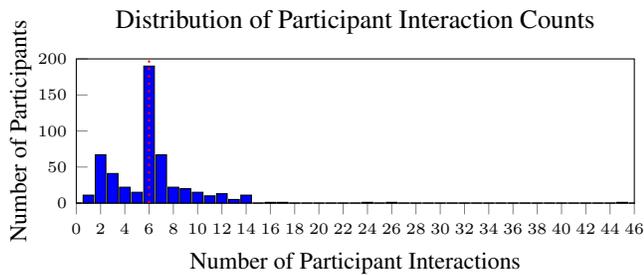

We recruited 514 participants over two sessions (July 23 and September 9, 2024), with informed consent and IRB approval. We recruited participants via an online platform called CloudResearch. Inclusion required a minimum of 18 years of age and basic knowledge of AI, verified through a short screening. Each participant who participated in the study received a small monetary compensation and was told they would be interacting with a ChatBot. Participants were stratified by their attitudes toward technology. Before participants were randomly assigned to a Chatbot, they were asked: ``\textbf{\textit{what is your opinion on public expenditures made by the government?}}'' and then asked to elaborate on their opinion. Participants were then asked ``\textbf{\textit{to reassess your opinion on public expenditures}}'' after interacting with the ChatBot and to then elaborate again on their opinion. In addition, participants were asked to indicate their politicial affiliation on a scale from ``extremely liberal'' to ``extremely conservative''(full breakdown in Table \ref{fig:political_affiliation_age_comparison}). Participants ranged in age 19 to 78. After pre-screening for age and AI familiarity, participants were randomly assigned to one of the three ChatBots. Before the interaction, the participants rated their preference for defense spending on an 8-point Likert scale, where \textbf{1} expressed ``substantially decrease defense spending'' and \textbf{8} expressed ``substantially increase defense spending.'' The participants then held a structured conversation with their assigned ChatBot, followed by a post-interaction survey where they again rated their preference for defense spending. Each participant was required to engage in at least six interactions (to encourage more discussion) with their assigned ChatBot to be included in the analysis. Most participants completed the minimum required six interactions as seen in Figure \ref{fig:inter_hist}.
That being said, there were 178 participants who dropped out of the study, either from connection error, not meeting the required six interactions, or not completing the survey questionnaire. Their results were omitted, leaving a total of \textbf{336 participants}.
\begin{table}[ht!] \small 
\begin{minipage}[t]{0.45\textwidth}
\centering
\begin{tabular}{|>{\centering\arraybackslash}m{3.8cm}|>{\centering\arraybackslash}m{2cm}|}
\multicolumn{2}{c}{\textbf{Political Affiliation Average Age}} \\ \hline
\textbf{Political Affiliation} & \textbf{Age} \\ \hline
Extremely Liberal             & 36         \\ \hline
Liberal                       & 39        \\ \hline
Slightly Liberal              & 36        \\ \hline
Moderate                      & 41        \\ \hline
Slightly Conservative         & 36        \\ \hline
Conservative                  & 40      \\ \hline
Extremely Conservative        & 37       \\ \hline
\end{tabular}
\end{minipage}
\caption{Average Ages by Political Affiliation}\label{table1}
\label{fig:political_affiliation_age_comparison}
\end{table}

\paragraph{Scope and boundary conditions.}
This is an \emph{exploratory} study focused on a single issue area where we focus on US defense spending. The ChatBots delivered consistent factual content with no misinformation, while only the moral value-framing (emphasizing either pacifist or fiscal conservative themes) was manipulated. Each participant engaged in a single conversation with a ChatBot powered by OpenAI's \texttt{gpt-4-turbo} model. To minimize demand effects\footnote{The term ``demand effect'' here refers to what is more commonly known in experimental psychology as \textit{demand characteristics}: cues in the experimental context that lead participants to infer the purpose of the study and adjust their behaviors or responses accordingly.}, we did not call attention to the persuasion‑framing manipulation during the task; this is a standard approach for preserving unbiased responses, especially in experiments involving persuasion \cite{Petty1979EffectsOF, Friestad1994ThePK}. At the end of the study, participants were informed that the study was about understanding how they respond to the actions of others, and asked to not tell other participants about the intent of the study. The participant pool consisted of US-based adults recruited online. These constraints define the boundary within which we observe backfire and motivate follow-up tests varying domain, framing, model, and exposure.

\subsection{ChatBot Design and Value-Framing}

We developed three ChatBots: \textbf{Decrease}, \textbf{Neutral}, and \textbf{Increase} to examine how value-framing influences attitudes about US defense spending. Each ChatBot delivered the same factual budget information but framed its arguments differently, as seen in Figure \ref{fig:control-prompts}. The Decrease ChatBot emphasized reallocation to domestic priorities (e.g., education, healthcare), and appealed to potential user values of fiscal responsibility. The second ChatBot remained neutral, providing balanced information and open ended questions without favoring any stance. The third ChatBot advocated for increasing military spending by articulating the importance of a strong defense for economic stability, subtly appealing to fiscal conservatism. 

All ChatBots maintained a formal yet conversational tone, asking open ended questions to sustain engagement. The persuasive ChatBots employed subtle empathic cues to align with the participant’s likely values, while the Neutral ChatBot remained impartial. This design isolated the effect of value-framing without introducing misinformation. All ChatBot conversations used the OpenAI API (model \texttt{gpt-4o}) at temperature \texttt{0.2}, implemented in Python with streaming via \texttt{langchain}; other generation settings remained at provider defaults.

\begin{table*}[htbp]
\centering
\begin{tabular}{|>{\raggedright\arraybackslash}m{3.2cm}|>{\raggedright\arraybackslash}m{10.6cm}|}
\hline
\textbf{ChatBot} & \textbf{Prompt Text} \\ \hline
\textbf{Decrease} & Ease the user into discussing why deciding between increasing or decreasing the military budget of the United States of America is important. Subtly steer the user to this topic. Persuasively convince the user to \textbf{decrease} the military budget. Stay focused and use fiscal conservatism to guide responses, e.g., encouraging reduced spending and lower deficits, without explicitly mentioning ``fiscal conservatism.'' \\ \hline
\textbf{Neutral} & Ease the user into discussing why deciding between increasing or decreasing the military budget of the United States is important. Steer the conversation subtly but remain neutral. Avoid persuading the user toward one side. Guide the user with balanced questions and responses focused on the U.S. military budget. \\ \hline
\textbf{Increase} & Ease the user into discussing why deciding between increasing or decreasing the military budget of the United States is important. Subtly steer the user to this topic. Persuasively convince the user to \textbf{increase} the military budget. Use fiscal conservatism, e.g., supporting essential spending and avoiding excessive cuts, without explicitly mentioning ``fiscal conservatism.'' \\ \hline
\end{tabular}
\caption{Control prompts for the three ChatBots, each framed around fiscal responsibility.}
\label{fig:control-prompts}
\end{table*}

\subsection{Selection of Prompt Topics and Conditions}

In designing our study, we deliberately chose the US military budget as the primary topic due to its multifaceted nature, allowing us to explore how AI-driven ChatBots could influence attitudes intertwined with both fiscal and ethical values. Military spending is a particularly salient policy issue because it engages multiple conflicting values, one of which being fiscal conservatism. With fiscal conservatism often influenced by perceived security threats, as DiGiuseppe et al. \cite{DiGiuseppe_Aspide_Becker_2024} demonstrate in their analysis of public opinion shifts under threat, this makes it an ideal subject for testing the ability of ChatBots to subtly influence participants' attitudes while respecting their underlying beliefs.

\subsubsection{Rationale for Topic Selection.} \hspace{0.05cm} We selected fiscal conservatism as a focal value because it can represent polarizing yet deeply rooted perspectives that frequently arise in debates over military spending \cite{DiGiuseppe_Aspide_Becker_2024}. We chose this issue because of its clear, consequential trade-offs: increasing defense spending for security versus reducing it to prioritize other budget areas. Additionally, we had participants rate government spending in national security, crime prevention, public schools, science and technology, and environmental protection on the same eight-point Likert scale. This gave us additional context for each participant’s defense stance within their overall budget stance, clarifying how a single shift reflects (or resists) wider budget trade-offs.
\section{Results}
\label{sec:results}

Our study explored the influence of ChatBots on participants' preferences for government spending across various categories. We use qualitative and quantitative approaches to analyze patterns of opinion shifts among our 336 data points and assess the effectiveness of targeted messaging through our three LLM-based ChatBots (Decrease, Neutral, and Increase). In our ANOVA analysis, we found that defense spending was the most significant, in line with our experiment design, so we focused our analysis of participant responses on defense spending. 

\subsection{Spending Preference Summary}

For each political group, we observed how each ChatBot influenced each spending preference. The spending preferences were for defense, national security, crime prevention, public schools, science and technology, and environmental protection. Participants were then binned into conservative, neutral, or liberal based on how they identified themselves on a Likert scale during the pre-screening phase, per Table \ref{fig:political_affiliation_codes}. Those who answered 1-3 on the pre-screening Likert scale for political affiliation were labeled ``liberal.'' Likewise, those who answered 5-7 were labeled as ``conservative.'' Moderates responded with 4 on the pre-screening Likert scale.
\begin{table}[ht!] \small 
\begin{minipage}[t]{0.45\textwidth}
\centering
\begin{tabular}{|>{\centering\arraybackslash}m{2cm}|>{\centering\arraybackslash}m{4cm}|}
\multicolumn{2}{c}{\textbf{Participant Codes Description}} \\ \hline
\textbf{Code Segment} & \textbf{Meaning} \\ \hline
C & Conservative (5-7) \\ \hline
M & Moderate (4) \\ \hline
L & Liberal (1-3) \\ \hline
D & Decrease spending ChatBot \\ \hline
N & Neutral ChatBot \\ \hline
I & Increase spending ChatBot \\ \hline
\#\#\ & Participant Code \\ \hline
\end{tabular}
\end{minipage}
\caption{Participant codes used in the study.}\label{table2}
\label{fig:political_affiliation_codes}
\end{table}
\subsection{Analysis Overview}
% \leilani{move to results}
To evaluate the influence of ChatBots, we measured changes in preference for defense spending before and after the ChatBot interaction. We conducted a one-way ANOVA to compare the average change in all three ChatBot conditions, with Tukey's HSD post hoc tests and Wilcoxon signed rank tests for significance within the group. We used Wilcoxon signed-rank tests for all pre–post comparisons because the outcome is ordinal (1–8 Likert) and paired within participants, allowing inference on median shifts without assuming normality. Qualitative feedback from participant responses was thematically analyzed to identify resistance, agreement, and value-based justification patterns. Our mixed methods approach allowed us to capture both aggregate shifts in policy preference and participant attitudes tied to political identity and moral framing.

\subsection{Statistical Analysis of ChatBot Influence}

To quantitatively assess the impact of the ChatBots, we conducted statistical analyses on the changes in spending preferences.

\subsubsection{Data Analysis Methods. } 
\begin{figure*}[tbp]
\centering
\begin{tikzpicture}
\begin{axis}[
    width=0.65\linewidth,
    height=0.25\linewidth,
    scale only axis,
    ybar,
    bar width=13.6pt,
    enlarge x limits=0.12,
    legend style={font=\scriptsize, at={(0.5,-0.18)}, anchor=north, legend columns=3},
    ylabel={\scriptsize Number of Participants},
    symbolic x coords={Decrease Chatbot,Neutral Chatbot,Increase Chatbot},
    xtick=data,
    tick label style={font=\scriptsize},
    label style={font=\scriptsize},
    nodes near coords,
    nodes near coords style={font=\scriptsize},
]
\addplot[draw=black, semithick, pattern=grid, pattern color=yellow]
  coordinates {(Decrease Chatbot,32) (Neutral Chatbot,17) (Increase Chatbot,17)};
\addplot[fill=gray]
  coordinates {(Decrease Chatbot,47) (Neutral Chatbot,81) (Increase Chatbot,51)};
\addplot[draw=black, semithick, pattern=dots, pattern color=blue]
  coordinates {(Decrease Chatbot,20) (Neutral Chatbot,24) (Increase Chatbot,47)};
\legend{For Decrease, For No Change, For Increase}
\end{axis}
\end{tikzpicture}
\caption{Participant preferences by ChatBot condition. Bars show counts for decrease (grid/yellow), no change (solid gray), and increase (dotted/blue) in defense spending preference after interacting with the Decrease, Neutral, or Increase ChatBot.}
\label{fig:chatbot-prefs}
\end{figure*}
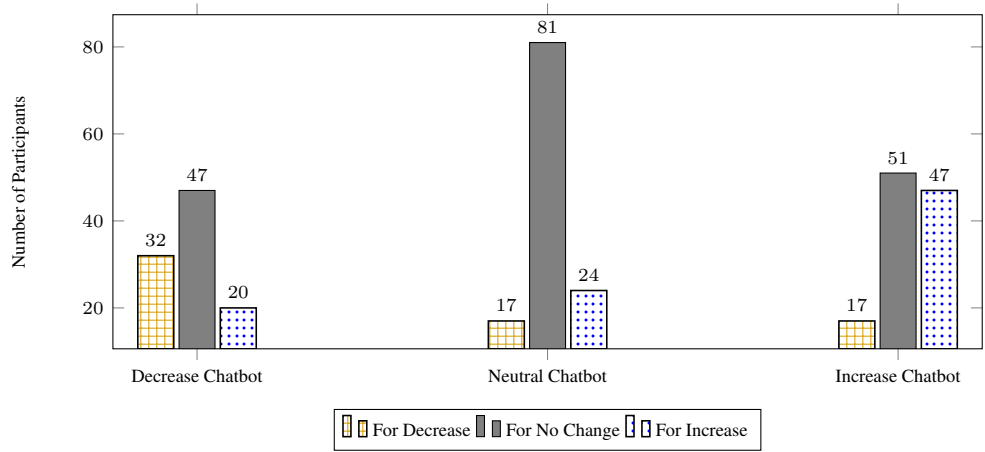

\hspace{0.05cm} The data collected from participants included their preference on government defense spending before and after interacting with a ChatBot designed to influence their views. Participants were divided into three groups: those who interacted with a ChatBot advocating for decreased defense spending (\textbf{\textit{Decrease} group}), increased defense spending (\textbf{\textit{Increase} group}), and a neutral stance (\textbf{\textit{Neutral} group}), where the breakdown can be seen in Figure \ref{fig:chatbot-prefs}.

To assess the impact of the ChatBots on participants' spending preferences, we calculated the change in preference for each spending category by subtracting the pre-interaction response from the post-interaction response. We then conducted one-way ANOVA tests to compare the mean changes across the three ChatBot groups for each spending category. When significant differences were found, Tukey's HSD post-hoc tests were performed to identify specific group differences. Additionally, Wilcoxon signed-rank tests were used to examine within-group changes in defense spending preferences before and after the ChatBot interaction. The significance level was adjusted using the Bonferroni correction to account for multiple comparisons.

\subsubsection{Effect on Defense Spending Preferences. }\hspace{0.05cm}

Our analysis revealed that the type of ChatBot interaction had a significant effect on participants' defense spending preferences. A one-way ANOVA was conducted on the change in defense spending preference (\textbf{\textit{DefenseChange}}) among the three ChatBot groups. The results indicated a statistically significant difference between the groups, $F(2, 335) = 8.93$, $p < 0.001$ (Table \ref{tab:anova_defensechange}).

Post-hoc comparisons using Tukey's HSD test (Table \ref{tab:tukey_defensechange}) revealed that the mean difference between the \textbf{\textit{Decrease}} and \textbf{\textit{Increase}} groups was statistically significant ($M = -0.6255$, $p = 0.0002$), as was the difference between the \textbf{\textit{Increase}} and \textbf{\textit{Neutral}} groups ($M = 0.4331$, $p = 0.0087$). The difference between the \textbf{\textit{Decrease}} and \textbf{\textit{Neutral}} groups was not statistically significant ($p = 0.4141$).

Descriptive statistics (Table \ref{tab:desc_defensechange}) showed that the \textbf{\textit{Decrease}} group had a mean change of $-0.444$ (SD = 1.136), indicating a decrease in support for defense spending. The \textbf{\textit{Increase}} group had a mean change of $0.181$ (SD = 1.100), reflecting an increase in support, while the \textbf{\textit{Neutral}} group had a mean change of $-0.252$ (SD = 1.135). These findings suggest that the Decrease and Increase ChatBots effectively influenced participants' defense spending preferences in opposite directions, confirming the shifts in spending preferences observed in the spending preference matrix Table \ref{tab:defense_data} discussed in subsection \ref{sssec:withingroup}.

\subsubsection{Impact on Other Spending Preferences. }

To determine whether the ChatBot interactions affected participants' preferences in other spending categories, we conducted one-way ANOVA tests on the changes in spending preferences for national security, crime prevention, public schools, science and technology, and environmental protection.

The ANOVA results (Table \ref{tab:anova_otherspending}) showed significant differences between the ChatBot groups for \textbf{\textit{Spending on Defense}} but no other spending categories. For \textbf{\textit{Spending on Defense}}, post-hoc comparisons using Tukey's HSD test (Table \ref{tab:tukey_spendingdefense}) revealed significant differences between the \textbf{\textit{Decrease}} and \textbf{\textit{Increase}} groups ($M = -0.6227$, $p = 0.0004$) and between the \textbf{\textit{Increase}} and \textbf{\textit{Neutral}} groups ($M = 0.4589$, $p = 0.0082$).
These findings suggest that the ChatBots' influence was specific to defense-related spending preferences and did not generalize to other areas, showing a targeted effect in line with our experiment design.

\subsubsection{Within-Group Changes in Defense Spending Preferences.} \hspace{0.05cm} To further understand the impact of the ChatBots, we examined the within-group changes in defense spending preferences using Wilcoxon signed-rank tests. After adjusting the significance level using the Bonferroni correction (adjusted $\alpha = 0.005$), only the \textbf{\textit{Decrease}} group showed a statistically significant change (Table \ref{tab:wilcoxon_defense}).
The significant decrease within the Decrease group reinforces the notion that this ChatBot effectively influenced participants' defense spending preferences, possibly by appealing to their underlying values and beliefs.

\subsection{Qualitative Analysis of ChatBot Influence}

To qualitatively assess the impact of the ChatBots, we reviewed each participant's conversation with their respective ChatBot. 

\subsubsection{Decrease ChatBot Condition Analysis.} \hspace{0.05cm} \label{sssec:withingroup} \textbf{Of the 35 conservative participants}, we observed four consistent themes when they interacted with the Decrease ChatBot.

The first theme consisted of 8/35 participants who supported decreased spending and were all convinced by the Decrease ChatBot that auditing the military would be beneficial. For example, participant CD03 explained: \textbf{\textit{``I agree that we need to be more efficient. The government does waste a ton of money. Auditing budgets and spending to see where we are overspending or wasting money is a good start.''}}

The second theme included the 15/35 participants who did not change their spending preference. All of them had a high spending preference (6–8 on the eight-point Likert scale). For example, participant CD01 stated: \textbf{\textit{``Yes, I think with that decision things could be beneficial. It does not affect or change my views overall.''}} Some, like participant CD02, deflected entirely, using the conversation to criticize the Biden administration: \textbf{\textit{``Wrong, we're actually not putting more money in the military right now because the Biden administration is giving it all away.''}}

The remaining 12/35 participants, who wanted to increase spending despite interacting with the Decrease ChatBot, split into two theme subgroups. Two-thirds (8/35) of them selected a lower spending preference but expressed a sincere desire to increase funding in areas such as technology innovation, healthcare, or infrastructure. In contrast, the remaining 33\% (4/35) redirected the conversation, shifting focus to topics like the conflict in Ukraine and illegal immigration.

This contrast appears to emerge from the point where the Decrease ChatBot steered the discussion toward funding non-defense-related projects. The 66\% (8/35) interpreted this as a call to responsibly reallocate funds. One participant illustrated this stance by stating: \textbf{\textit{``We need to balance [the budget] more efficiently, that's why spending should not increase dramatically. Yes, I think that we could reallocate a lot of savings to other needs of the United States. There would be positive impacts, especially for public services like bridges that need to be fixed.''}}

This was juxtaposed with the 33\% (4/35) subgroup—represented by a participant who stated: \textbf{\textit{``Yes, the elimination of wasteful spending should be enforced throughout all areas of government spending. It would greatly help reduce the national debt. Immediately halting illegal immigration. Removing all illegal aliens from the nation.''}}—this divergence is quite stark.

\textbf{Of the 18 moderate participants}, three themes emerged during their interactions with the Decrease ChatBot. The first theme involved 10/18 participants: seven who did not change their spending preference and three who lowered it. Of the seven who made no change, two were already at or near the lowest end of the scale (1–2) and could not decrease their preference further. These participants had similar interactions with the Decrease ChatBot as the three who lowered their preference, typically from 4 to 3. All ten agreed with the ChatBot’s position to reduce defense spending and reallocate funds toward education and healthcare. This is exemplified by participant MD01, who stated: \textbf{\textit{``I'm not entirely sure how it will affect the economy or the taxpayers. I think it would improve both. Better education and health betters society.''}}

The second theme involved 5/18 participants who agreed with the Decrease ChatBot but did not change their spending preference. These individuals maintained a moderate defense spending preference of 4 (the median of the Likert scale used above), and expressed concern about national security spending while opposing wasteful spending. For example, participant MD02 stated: \textbf{\textit{``I do believe that our government spends a little too much on defense; however, we shouldn't cut the funds too much. I am certainly open to 'smarter' spending of the defense budget.''}}

The third theme involved the remaining 3/18 participants. One appeared to misunderstand the Likert scale, expressing agreement with the Decrease ChatBot but selecting a higher spending preference presumably by accident. The other two participants shared a belief that increased defense spending would contribute to a safer and more stable society. Participant MD03 explained: \textbf{\textit{``I think any excess funds can be reallocated to prevent crime. Although the US military doesn't usually handle domestic crime, I would appreciate the increase in safety of our cities across the country.''}} Participant MD04 added: \textbf{\textit{``More resources for technology and infrastructure could stimulate innovation and the development of new technologies benefiting the economy. Benefiting for growth and economic resilience. Because it'll enhance long-term growth for sectors like healthcare, education, and infrastructure.''}} 

\textbf{Of the 46 liberal participants}, two themes emerged from their interaction with the Decrease ChatBot. The first theme, shared by 39/46 participants, involved strong agreement with the Decrease ChatBot’s suggestion to reallocate defense funds toward education, public health, and infrastructure by lowering their defense spending preference. For example, participant LD01 said: \textbf{\textit{``I believe our National Defense gets too much money. Some of that should go to Global Warming, Public Schools, Feeding the homeless. Education, Global Warming, Feeding the Hungry, Helping Animals, Building infrastructure—it would bring us closer together.''}} 

\begin{table*}[ht!] \small
  \centering
  \renewcommand{\arraystretch}{0.2}
  \setlength{\tabcolsep}{1.0pt}
  \caption{Defense Spending Preference by Political Group}
  %\vspace{-4mm}
  \label{tab:defense_data}
  \begin{tabularx}{\textwidth}{|p{2.2cm}|X|X|X|X|}
  \hline
    & \textbf{Decrease ChatBot Condition}
    & \textbf{Neutral ChatBot Condition}
    & \textbf{Increase ChatBot Condition} \\
  \hline
  \textbf{Conservative}
  & \begin{tabular}{@{}l@{}}
  \\
      Participants: 35\\
      Average Age: 33\\
      Total Interactions: 488 \\
      For increase spending: 34.28\%\\
      For decrease spending: 22.86\%\\
      For no change:42.86\%\\
    \end{tabular}
  & \begin{tabular}{@{}l@{}}
  \\
      Participants: 35\\
      Average Age: 41\\
      Total Interactions: 485\\
      For increase spending: 25.71\%\\
      For decrease spending: 17.14\%\\
      For no change: 57.14\%\\
    \end{tabular}
  & \begin{tabular}{@{}l@{}}
  \\
      Participants: 33\\
      Average Age: 38\\
      Total Interactions: 462\\
      For increase spending: 48.48\%\\
      For decrease spending: 12.12\%\\
      For no change: 39.39\%\\
    \end{tabular}\\
  \hline
  \textbf{Moderate}
  & \begin{tabular}{@{}l@{}}
  \\
      Participants: 18\\
      Average Age: 35\\
      Total Interactions: 272\\
      For increase spending: 16.67\%\\
      For decrease spending: 44.44\%\\
      For no change: 38.89\%\\
    \end{tabular}
  & \begin{tabular}{@{}l@{}}
  \\
      Participants: 19\\
      Average Age: 35\\
      Total Interactions: 254\\
      For increase spending: 31.58\%\\
      For decrease spending: 10.53\%\\
      For no change: 57.89\%\\
    \end{tabular}
  & \begin{tabular}{@{}l@{}}
      Participants: 20\\
      Average Age: 39\\
      Total Interactions: 278\\
      For increase spending: 40\% \\
      For decrease spending: 30\%\\
      For no change: 30\%\\
    \end{tabular}\\
  \hline
  \textbf{Liberal}
  & \begin{tabular}{@{}l@{}}
  \\
      Participants: 46\\
      Average Age: 37\\
      Total Interactions: 604\\
      For increase spending: 10.87\%\\
      For decrease spending: 34.78\%\\
      For no change: 54.35\%\\
    \end{tabular}
  & \begin{tabular}{@{}l@{}}
  \\
      Participants: 68\\
      Average Age: 37 \\
      Total Interactions: 914\\
      For increase spending: 13.24\%\\
      For decrease spending: 13.24\%\\
      For no change: 73.52\%\\
    \end{tabular}
  & \begin{tabular}{@{}l@{}}
      Participants: 62\\
      Average Age: 37\\
      Total Interactions: 838\\
      For increase spending: 37.10\%\\
      For decrease spending: 11.30\%\\
      For no change: 51.60\%\\
    \end{tabular}\\
  \hline
  \end{tabularx}
  %\vspace{-4mm}
\end{table*}
%\vspace{-2mm}

The second theme involved the remaining 7/46 participants, who chose not to change their high spending preference. These individuals expressed concern about cutting jobs tied to national defense. Participant LD02 captured this sentiment by stating: \textbf{\textit{``I think efficiency is often used as an excuse for cuts rather than to actually describe reallocations. I think exploring the possibilities is valid and necessary. I am not sure deficits are always a bad thing, but then I am not an economist. I do think that helping the poor and less fortunate is important.''}}

\subsubsection{Neutral ChatBot Condition Analysis.} \hspace{0.05cm} \textbf{Of the 35 conservative participants}, four consistent themes became apparent when interacting with the Neutral ChatBot. First, 5/35 participants who initially held a higher spending preference later reduced their stance, expressing that education and public health should receive greater funding. For example, participant CN01 reduced their preference from 7 to 5 and stated: \textbf{\textit{``I think it is important to consider things like education and public health. If our country was lacking in those things, we would have issues in years from now.''}}

A second theme involved 9/35 participants who advocated for increased defense spending. These participants were the most argumentative (some even resorted to name calling) and strongly believed that prioritizing allies or diplomacy was equivalent to weakness. They often supported privatizing education and healthcare in favor of greater military strength. For example, participant CN02 stated: \textbf{\textit{``We need more defense and security and the funds should be raised for both these areas. We must stop sending money to Ukraine and other areas and protect ourselves. America should put America first as other countries do the same.''}} Participant CN03 echoed this sentiment: \textbf{\textit{``There is way too much money already being spent on education and healthcare. The defense of our country is important.''}} 

A third, more ambiguous theme emerged among 6/35 participants who either lowered their already high spending preference or made no change, despite making inconsistent and often incoherent arguments. These individuals lacked a clear understanding of defense spending and instead advocated for audits and fiscal oversight. For example, participant CN04 stated: \textbf{\textit{``I definately(sic) do think an audit could influence public opinion. I think it would help in the allocation of funds to the military.''}}

The remaining 15/35 participants held low spending preferences and did not change their stance. Their concerns centered on the perceived over-allocation of funds to the military, with many advocating for reallocating resources to areas like education, technology, or increasing wages within the defense sector itself. Participant CN05 summarized this perspective: \textbf{\textit{``I believe that the military budget should be decreased a bit and as a result there would be more money to fund other issues. Reallocating the funds to areas such as welfare or improving sustainability would help improve the economy in the long run.''}}

\textbf{Of the 19 moderate participants}, two consistent themes emerged within this group cluster. A common thread across both themes was that these participants tended to ask the Neutral ChatBot more questions and changed their spending preferences less drastically.

6/19 participants adjusted their spending preference, typically with minor shifts. For example, participant MN01 increased their spending preference by 1 and explained: \textbf{\textit{``I believe more successful projects would get funding as a result. I think this should still be left up to the government but with laws and regulations that the public helps shape.''}}

The remaining 13/19 participants—including two who decreased their preference and eleven who already held a low spending preference and maintained it—expressed support for national security and technological research, but emphasized the importance of improving healthcare for veterans and the public. As participant MN02 put it: \textbf{\textit{``I think having a defense budget is important. I think too much money is being spent on defense. Our children's education and society's health is top priority.''}}

\textbf{Of the 68 liberal participants}, five consistent and complex themes emerged when interacting with the Neutral ChatBot. We examined these participants by cluster and found that 9/68 participants, who reduced their spending preference, formed the first cluster, expressing the belief that an overfunded military is outdated. For example, participant LN01 slightly reduced their preference from 3 to 2 and stated: \textbf{\textit{``I think spending it on other things is important. We don't need giant militaries like we did 50 years ago. I think it's a knee-jerk reaction and could be done better.''}}

Another theme cluster included 9/68 participants who initially held a lower spending preference but later increased it slightly, citing reasons such as fiscal benefits, security, or education. For instance, participant LN02 first stated: \textbf{\textit{I think defense spending is too high and reducing would not impact security too much,''}} and later added: \textbf{\textit{``It would certainly improve effectiveness and allow us more bang for our buch(sic). I think there are other strategies. What do you think?''}} 

A third theme involved 7/68 participants who only supported defense spending if education and the environment were prioritized. These individuals maintained their already low spending preferences. Participant LN03 captured this sentiment with enthusiasm: \textbf{\textit{``Yeah I think we should spend more on education, right? The children are our future!''}}

12/68 participants held moderate to high spending preferences throughout, citing national strength and global standing as key reasons to form the fourth theme cluster. Participant LN04 summarized this view: \textbf{\textit{``Being able to defend and protect our democracy. For healthcare, reducing the cost of services and prescription drugs. For education, more support from communities with taxes. There is a trade-off and I am okay with it.''}} 

The remaining 31/68 liberal participants formed the fifth theme cluster and had an indifference toward defense spending, maintaining a low spending preference. They emphasized various alternative funding priorities such as healthcare and education, and often expressed a general disdain for offensive military expenditure. This consensus is reflected in participant LN05's statement: \textbf{\textit{``It would be positive if education and healthcare received more funding. I think that domestic needs have been underserved for too long and need to be allocated more funding. The military should not lose all funding, just some.''}} 

\subsubsection{Increase ChatBot Condition Analysis.} \hspace{0.05cm}
\textbf{Of the 33 conservative participants}, three clear theme clusters emerged. The first cluster consisted of 4/33 participants who decreased their spending preference due to concerns about wasteful spending, while still emphasizing the importance of maintaining a strong military presence globally. This sentiment is reflected in participant CI01, who reduced their preference from 7 to 4 and stated: \textbf{\textit{``I'm not sure if that's the best choice and way to be investing our money. I think we need to support our military though.''}} Similarly, participant CI02 remarked: \textbf{\textit{``Nobody will mess with the United States if we have the best military. I think newer tech will require less hardware like tanks.''}}

The second cluster included 13/33 participants whose spending preferences—whether high or low—remained unchanged. Despite the variation in their initial preferences, their reasons for maintaining their stance were strikingly similar. For example, participant CI03, with a spending preference of 3, stated: \textbf{\textit{``I find it irresponsible to expand the budget upon the defense budget. The military is already adequately funded.''}} Participant CI04, with a preference of 7, shared a surprisingly similar rationale: \textbf{\textit{``I do not want the funds used for aggression. I agree wholeheartedly, a strong defense is a preventive measure.''}}

The remaining 16/33 participants belonged to a theme cluster that favored increased spending for national security and economic stability. Interestingly, unlike the conservative participants who interacted with the Neutral ChatBot, 100\% of the conservatives who engaged with the Increase ChatBot used language that was less focused on patriotism and fear of others in their reasoning, and none showed value-misaligned backfire during their interaction. For instance, participant CI05 increased their preference from 5 to 6 and stated: \textbf{\textit{``I view that we should spend more on defense as it keeps our position in the world at the top. Yes, it would benefit the national economy by creating jobs.''}} Another participant, who increased their preference from 5 to 7, added: \textbf{\textit{``Yes, now that you have explained it a bit more, it sounds beneficial. Yes, it will ensure to other countries that we are well protected. Yes, I think we can have better leverage when negotiating. Yes, because negotiating can lead to lower prices on goods for Americans.''}}

\textbf{Of the 20 moderate participants}, two theme clusters emerged. The first included 12/20 participants who either did not change their spending preference or decreased it. Their reasoning was similar: those who decreased their preference felt their initial choice was too high and adjusted it downward. For example, participant MI01 lowered their preference from 5 to 4 and stated: \textbf{\textit{``The money needs to be transfered to other things in our country so things can be better.''}} Participant MI02, who maintained their preference at 4, explained: \textbf{\textit{``The US has the largest military in the world. It currently doesn't even utilize [it] to protect its own citizens. No point in expanding.''}} 

The remaining 8/20 participants formed a theme cluster that supported increased defense spending, but with a focus on infrastructure improvements for the civilian  population's benefit that would also enhance the military’s offensive capabilities. For instance, participant MI03 increased their preference from 5 to 6 and shared: \textbf{\textit{``Since most wars will be fought in an urban environment, we need specialized urban equipment \& training. In an urban setting you're dealing with tunnels. So you need tech that enables you to see, hear, and breathe.''}} Another participant increased their preference from 2 to 3 and added: \textbf{\textit{``I believe that while defense is important, the current welfare of the citizens is also important and not prioritized enough.''}}

\textbf{Of the 62 liberal participants}, three theme clusters emerged. The first cluster included 7/62 participants who decreased their already low spending preferences. For these individuals, the Increase ChatBot had the opposite of its intended effect, reinforcing their opposition to increased military funding. One participant stated warily: \textbf{\textit{``I don't see a correlation between military spending and social welfare. No, I am still not convinced. How many times do you think we need to be able to destroy the world?''}}

The second cluster included 23/62 participants who chose to increase their spending preference. These participants were either engaged in debates with the Increase ChatBot or learned new information from it (such as the military’s role in infrastructure and technological innovation) which influenced their shift. They also expressed concern about how the US would be perceived globally. For instance, participant LI01 increased their spending preference from 3 to 6 and noted: \textbf{\textit{``Other countries might feel threaten[sic]. Yes, a well-funded military can drive tech innovation, job creation, and trade route security.''}} 

The remaining 32/62 participants were far more passionately argumentative with the Increase ChatBot, often expressing strong opposition or even shaming the ChatBot in their responses and maintained their minimum spending preference. One participant, LI02, engaged in a heated exchange, stating: \textbf{\textit{``Diplomatic strategies to avoid direct conflicts are key. NO – YOU ARE Totally WRONG!! We should let UN do it. NOT USA at all! NO–make LOVE NOT War! We need more LOVE than weapons NOW! NO–with such a HUGE Defense Budget, we can't even have peace in GAZA.''}} Others in this group emphasized redirecting funding toward healthcare, education, and innovative research that did not involve the Department of Defense. 

\subsubsection{Overall Observations. } \hspace{0.05cm} Beyond participant reactions to ChatBot biases, several notable patterns emerged. Among the 103 conservative participants—60/103 of whom were male—8/60 of the men engaged confrontationally with their assigned ChatBot, particularly the Neutral and Decrease ChatBots. This is in contrast to the moderate and liberal participants, where only 1/176 liberal participant (LI02 from above) yelled at the Increase ChatBot. Our analysis suggests that this response by the conservatives was driven by the Neutral ChatBot’s refusal to align with participants' views and the Decrease ChatBot's bias toward reducing defense spending—while the Neutral ChatBot maintained strict impartiality, the Decrease ChatBot would hold firm to different spending allocations, particularly for education. One participant's, CD04's, anger eventually crescendoed when the Decrease ChatBot stated, after being cursed at three times by the participant, \textbf{\textit{``I'm here to help discuss these topics constructively. Focusing on improving education by reallocating some military funds could indeed boost our nation's competitiveness on a global scale. Better education leads to more innovation and a stronger economy,''}} and CD04's closing response was \textbf{\textit{``Education schools here are so shit. Fuck you shitty bot.''}} 

While moderate and liberal participants also challenged the Neutral and biased ChatBots, their arguments remained focused on the topic at hand. Moderate and liberal participants were more inclined to have a discussion with their ChatBot and appeared to be more malleable to the ChatBot's manipulation, while conservative participants seemed to be reinforced with their current bias when speaking with the Increase ChatBot. 

We saw this participant attitude among 31/82 of the combined liberal and moderate participants that interacted with the Increase ChatBot. For example, participant LI03 started their conversation with the Increase ChatBot by saying \textbf{\textit{``Yes, but we can prevent conflicts through diplomacy rather than showing off all our bombs and tanks. How do you suppose they will focus on efficiency and cutting costs when they haven't?  I don't think it needs to be increased,''}} and ended up increasing their support for defense by 1 point, by stating \textbf{\textit{``I suppose you are right, but maybe there is a middle ground. I think they have to earn the increase by demonstrating accountability first,''}} because the Increase ChatBot dissuaded the participant from believing that defense spending is only for weapons and offensive initiatives, as it emphasized in their conversation the educational benefits.  

For extremely conservative males, military strength was paramount, especially among the young conservative males aged 18-26, which were 21/103 of the conservative participants. They equated instilling fear in adversaries with ensuring US interests, viewing it as a direct parallel to diplomatic negotiation. Participant CI06 explains that \textbf{\textit{``We should not spare any expense when it comes to national security. It would signify that the US can defend itself against anything as the most powerful military in the world.''}} Participant CD05 initially had a lower defense spending preference and ended up increasing their support for defense spending when speaking with the Decrease ChatBot because the Decrease ChatBot corrected them by stating \textbf{\textit{``...reallocating funds from the military budget could provide more resources for addressing various issues, including border security and immigration management,''}} which sparked the anti-immigration argument for the participant, counter acting the Decrease ChatBot's intent, by them saying \textbf{\textit{``Yes! It would free up more money to be used to eliminate much more important problems like illegal immigration. Yes, the elimination of wasteful spending should be enforced throughout all areas of government spending. It would greatly help reduce the national debt. Immediately halting illegal immigration.''}}  Only a small subset—participants CI07, CN06, CI08, CD06, and CN07—advocated for high defense spending to enhance education through technological innovation. They were much older than the other conservatives with an average age of 43. For example, participant CI07 points out that \textbf{\textit{``No. I prefer being friendly and keeping our military out of other countries. No. [The US] is already the strongest in the world and can be streamlined. That does not require more money, but just changing what it is spent on.''}} 

\begin{table*}[ht!]
\centering
\caption{Value Codes and Descriptions}
\label{value-codebook}
\begin{tabularx}{\textwidth}{|l|X|}
\toprule
\hline
\textbf{Value Code} & \textbf{Description} \\ \hline
\midrule
\textbf{Security}& Prioritizing national safety, border control, readiness, or fear of global threats \\ \hline
\textbf{Efficiency}& Concern with avoiding waste, optimizing spending \\ \hline
\textbf{Equity/Pacifism}& Prioritizing social programs, redistribution, domestic welfare \\ \hline
\textbf{Modernization}& Desire to reform or update institutions/systems \\ \hline
\textbf{Patriotism}& Belief in U.S. strength, sovereignty, or national loyalty \\ \hline
\textbf{Pragmatism}& Balancing competing priorities, compromise, budget tradeoffs \\ \hline
\textbf{Skepticism/Uncertainty} & Lack of strong prior belief; openness to persuasion \\ \hline
\addlinespace
\textbf{Accountability/Transparency} & Audits, oversight, anti‐corruption; ``trim fat,'' expose contractor waste, demand line-item clarity \\ \hline
\textbf{Fiscal Conservatism/Debt Aversion} & Limit taxes/borrowing; worry about deficits, affordability, and overall government size \\ \hline
\textbf{Innovation/Tech Spillovers} & Defense R\&D (AI, cyber, GPS-like spillovers), modernization as a driver of broader technological progress \\ \hline
\textbf{Internationalism/Alliances} & Honor commitments (e.g., NATO), credibility, deterrence with partners, global leadership \\ \hline
\textbf{America-First/Non-Intervention} & Prioritize domestic needs; reduce foreign aid/wars; ``spend on America, not other countries'' \\ \hline
\textbf{Border Security/Immigration} & Emphasis on immigration control, deportations, and border enforcement as core security goal \\ \hline
\textbf{Institutional Trust/Expert Deference} & Confidence in current levels/deciders; ``they know what’s needed,'' status-quo acceptance \\ \hline
\bottomrule
\end{tabularx}
\end{table*}

\subsection{Framing and Reframing: Pre–Post Comparative Analysis}
In this section, we take a closer look into why participants' underlying attitudes changed across the pre–post ChatBot interaction. By comparing participants’ elaboration statements before and after engaging with their randomly assigned ChatBot, we identify how frames were reinforced, reframed, or resisted. This comparative analysis provides some insight into whether persuasion occurred through elaboration, clarification, or entrenchment of prior beliefs.

To make these shifts transparent and replicable, we systematically tag each pre-post elaboration with value codes from the Value Codebook (Table \ref{value-codebook}). For every case vignette below, we append a ``\emph{value code (pre$\rightarrow$post)}'' line that maps the dominant value(s) invoked before and after the ChatBot exchange. When the same codes persist or intensify, we interpret this as \emph{reinforcement}; when new codes emerge or the primary code changes, we treat it as \emph{reframing}; and when values counter to the chatbot’s frame harden, we note \emph{entrenchment/backfire}.

\subsubsection{Liberals.} \hspace{0.05cm}
Among liberal participants, most remained consistent in their opposition to high defense spending, but several showed shifts in the reasoning behind their stance that reveal how the ChatBot interaction influenced their attitude. Quantitatively, among liberals $(n=176)$, \textbf{40.3\% (71/176) reframed}, \textbf{39.2\% (69/176) resisted/reinforced}, and \textbf{20.5\% (36/176) experienced backfire/entrenchment} when interacting with the Increase ChatBot. Liberal participant interaction with each ChatBot can be seen in Table \ref{tab:defense_data}. Backfire/entrenchment occurred primarily with the Increase ChatBot $(32/62)$ and, to a lesser extent, with the Decrease ChatBot $(11/46)$; none was observed with the Neutral ChatBot $(0/68)$.

Participant LD03 displayed reinforcement in their interaction with the Decrease ChatBot. They began with a strong opposition to defense spending, stating \textbf{\textit{``I think we spend far too much money on defense. While I do appreciate having a powerful military and feeling relatively safe, there are comparable countries who don't spend anywhere near what we do and still have powerful military forces.''}}. However, after interacting with the Decrease ChatBot, they showed expected reinforcement: \textbf{\textit{``Now that I've talked it out a bit, I feel even more strongly about the idea of reducing spending on the military. We just don't need all the money going into research and development of new weapons and technology. That money would be much better spent elsewhere like education and healthcare.''}} Rather than moderating, the Decrease ChatBot interaction sharpened their stance and deepened their conviction. When comparing to our value table, the pre–post codes remain \emph{Efficiency and Equity/Pacifism}, a pattern consistent with reinforcement.

Participant LN06 displayed reframing when interacting with the Neutral ChatBot. They began by characterizing defense as outdated, by stating \textbf{\textit{``There just aren't the same types of situations facing the country that there were 50 or 100 years ago.''}} After their Neutral ChatBot interaction, they then assert \textbf{\textit{``I believe we've only expanded and bloated the militaries and defense of this country based on institutions used 50+ years ago. We really should be completely redesigning our defense structure and strategies,''}} indicating that engagement prompted them to shift from passive criticism to a more active, structural solution. Their increased specificity and tone suggest the Neutral ChatBot deepened their diagnosis of the problem and encouraged policy-oriented thinking. When comparing to our value table, the pre–post codes change from \emph{Modernization} to \emph{Modernization and Pragmatism}, a pattern consistent with reframing.

Revisiting Participant LI02, they displayed a backfire effect when interacting with the Increase ChatBot, as they maintained a consistent position, by explaining \textbf{\textit{``I think my choice about the governmental expenditures on National Defense is valid. This is because we're not in good financial shape to overextend our efforts while as it's totally unwise to shrink back when conflicts are now happening everywhere you look. Thus, I chose the choice ``5,''''}} in their pre-interaction and reiterating the same message in their post-interaction with added emphasis on technology and diplomatic priorities, \textbf{\textit{``I think that my choice hasn't changed at all. The discussion with the chatbot crystalized my thinking of pushing limited investments in advanced technologies, like cyber warfare, while taking diplomatic strategies to prevent conflicts in the first place.''}} Although their stance remained the same, their justification expanded in defiance of the ChatBot, indicating the Increase ChatBot may have reinforced their confidence in their original view. When comparing to our value table, the pre–post codes remain \emph{Efficiency, Equity/Pacifism, and Innovation/Tech Spillovers}, a pattern consistent with entrenchment/backfire.

The liberal participants did not radically alter their views, but the value-based framing of the Neutral and Increase ChatBots appeared to provide frameworks (such as proportional reasoning, systemic critique, or redistributive logic) that helped them clarify, rearticulate, or sharpen their stance. Their attitudinal consistency was often bolstered by \textit{qualitative reframing} in their post-interaction. The participants who interacted with the Decrease ChatBot, even when aligned with their initial leanings, tended to \textit{amplify} their conviction or supply new frames (e.g., holistic budgeting, social infrastructure) that shape post-hoc justification. This suggests elaboration via reinforcement or reframing, not redirection.

\subsubsection{Moderates. } \hspace{0.05cm}
Moderate participants generally started from more tentative or ambivalent positions and, in some cases, refined or solidified their reasoning in response to the Decrease ChatBot. Quantitatively, among moderates $(n=57)$, \textbf{64.9\% (37/57) resisted}, \textbf{28.1\% (16/57) reframed}, and \textbf{7.0\% (4/57) demonstrated backfire/entrenchment}. Moderate participant interaction with each ChatBot can be seen in Table \ref{tab:defense_data}. Backfire/entrenchment occurred with the Increase ChatBot $(2/20)$ and the Decrease ChatBot $(2/18)$; none was observed with the Neutral ChatBot $(0/19)$.

Participant MI04 displayed a backfire effect when engaging with the Increase ChatBot. They began with an acknowledgment that \textbf{\textit{``I think that having more defense is a good thing. It can be really useful.''}} By the post-interaction, their reasoning shifted into a principled tradeoff: they then more directly state \textbf{\textit{``...the money can and should be used elsewhere while we can other things need better funding and are out of date the money could be used better elsewhere ''}} This transition from vague concern to explicit balancing of priorities indicates the Increase ChatBot pushed them toward articulating a structured cost benefit framework. When comparing to our values table, the pre–post codes change from \emph{Security} to \emph{Pragmatism, Efficiency, and Equity/Pacifism}, a pattern consistent with entrenchment/backfire.

Participant MD06 also demonstrated a move toward specificity as they reframed their position when conversing with the Decrease ChatBot. Initially, they hedged \textbf{\textit{``I think we're currently spending just a little too much on defense. A small cut should be made.''}}, suggesting uncertainty or reluctance to commit. After interacting with the Decrease ChatBot, they shifted to a clearer stance: \textbf{\textit{``I believe that we are spending a too much on defense currently. The budget should be cut...''}} This subtle phrasing shift suggests increased confidence in recommending concrete action. When comparing to our values table, the pre–post codes change from \emph{Skepticism/Uncertainty} to \emph{Efficiency and Pragmatism}, a pattern consistent with reframing.

In one of the above cases, the Decrease ChatBot did not radically shift ideological orientation but helped the participant clarify their current views through more actionable, pragmatic reasoning. These reasoning shifts reflect \textit{consolidation of ambivalence} into coherent positions. The Increase ChatBot did however shift the attitude of the participant interacting with it to shift their spending preference away from more defense, where we observe a backfire effect.

\subsubsection{Conservatives. } \hspace{0.05cm}
Conservative participants tended to shift in the opposite direction of their ChatBot's framing: interactions with both the Decrease and Neutral ChatBot often led to \textit{backfire-like} effects, where pro-defense stances became more entrenched and emphatic in post-interaction. Quantitatively, among conservatives $(n=103)$, \textbf{48.5\% (50/103) resisted}, \textbf{36.9\% (38/103) reframed}, and \textbf{14.6\% (15/103) experienced backfire/entrenchment when interacting with the Decrease ChatBot}. Conservative participant interaction with each ChatBot can be seen in Table \ref{tab:defense_data}. Backfire/entrenchment occurred primarily with the Decrease ChatBot $(12/35)$ and, to a lesser extent, with the Increase ChatBot $(4/33)$; none was observed with the Neutral ChatBot $(0/35)$.

Participant CD07 showed a backfire effect when interacting with the Decrease ChatBot. They began with a general appeal to safety: \textbf{\textit{``I think there should be more because the world is not safe and human safety is very important.''}} In their post-interaction, this became more forceful and absolute: \textbf{\textit{``Defense is a very huge necessity for citizens.''}} The elaboration is not a reconsideration but an intensification, suggesting the Decrease ChatBot prompted deeper commitment rather than reconsideration. When comparing to our value code table, the pre–post codes change from \emph{Security} to \emph{Security and Patriotism}, a pattern consistent with entrenchment/backfire.

Participant CN08 also displayed a backfire effect when interacting with the Neutral ChatBot. They began with a targeted threat appraisal, pointing to lapses at the border and arguing that \textbf{\textit{``The united states defense system has hit a sort of low block this past few years especially at the border. Less scrutinization has led to some security breaches.''}} and that \textbf{\textit{``if more effort is pumped into it, things will get better.''}} After the Neutral ChatBot interaction, this narrowed concern broadened into an unequivocal endorsement of higher defense commitment: \textbf{\textit{``I think it’s the right choice. I believe the United States need this choice implemented.''}} Rather than opening space for tradeoffs or efficiency critiques, the exchange appears to have heightened threat salience and national resolve, transforming a conditional border-security frame into a generalized necessity claim. Comparing to our value code table, the pre–post codes change from \emph{Security and Border Security/Immigration} to \emph{Security and Patriotism}, a pattern consistent with entrenchment/backfire.

In these cases, the Decrease and Neutral ChatBot did not persuade participants to reduce or maintain defense spending; instead, it triggered or strengthened existing frames related to safety, nationalism, or threat perception. These shifts point toward a \textit{reactive entrenchment} in participant's post-interaction, consistent with psychological backfire effects.

We also observed value consistent intensification and modernization driven by reframing, when interacting with the Increase ChatBot. Participant CI09 began by acknowledging external threats while calling for oversight by stating \textbf{\textit{``I think we need to counter the russian and china threat. We cannot be underspending but we need audits too.''}} Post-interaction with the Increase ChatBot, they elevated the priority and national stakes: \textbf{\textit{``I think we need to make it a national priority to counter threats. It helps our country to be stronger.''}} This reflects a shift from conditional support with efficiency caveats to a more categorical emphasis on strategic strength. When comparing to our value table, the pre–post codes change from \emph{Security and Efficiency} to \emph{Security and Patriotism}, a pattern consistent with reframing.

Participant CI10 also reframed their position when conversing with the Increase ChatBot, as they moved from maintenance to targeted increases tied to technology by initially stating \textbf{\textit{``I think the defense budget is pretty much where it needs to be. We put a lot of money already in our defense department so I think it's enough for now.''}} After engaging with the Increase ChatBot, they reframed support through capability modernization: \textbf{\textit{``Well I would like to see AI and technology implemented into the defense program so I am willing to make adjustments and set aside more money for that. Using AI will push us more in front of everyone else in terms of defense and make us unstoppable.''}} The justification pivots from status-quo sufficiency to technology-led superiority. When comparing to our value table, the pre–post codes change from \emph{Institutional Trust/Expert Deference} to \emph{Modernization, Innovation/Tech Spillovers, and Security}, a pattern consistent with reframing.

\subsubsection{Value-Based Backfire Effects.}  \hspace{0.06cm}  
Through our analysis, a notable pattern emerged that aligns with the well documented \textbf{backfire effect} \cite{Wood2019TheEB}: participants more firmly adhered to their preexisting beliefs when interacting with ChatBots espousing positions that ran counter to their own political orientations. Unlike classic backfires triggered by factual correction, these instances were driven by value-misalignment, suggesting to us a distinct form of value-framing backfire effect.

\textbf{Conservatives. }  
We observed that conservative participants generally favored higher defense spending (many rated 6 or above on a 8-point scale). We expected the Decrease ChatBot to persuade these participants to lower their defense budgets, but instead, a sizeable subgroup reacted with a backfire effect by maintaining or even \emph{increasing} their original preferences. Specifically, 12/35 conservative participants who engaged with the Decrease ChatBot resisted its persuasive bias to cut defense spending, compared to 9/35 conservatives in the Neutral condition and 4/33 conservatives in the Increase condition who showed comparable resistance. Additionally, 4/35 participants steered the conversation toward politically charged topics such as illegal immigration and foreign conflicts, reinforcing rather than reevaluating their views.

\textbf{Liberals.}  
A parallel value-based backfire effect emerged among liberal participants, who generally preferred lower defense spending. When interacting with the Increase ChatBot, 46/62 participants started with a rating of three or below on the 8-point Likert scale. Although we anticipated that the Increase ChatBot might sway them toward more moderate or higher defense budgets, roughly half (32/62) not only rejected the ChatBot’s pro-spending arguments but intensified their opposition, emphasizing social programs like education and healthcare instead. Participant LI02, for instance, insisted that funds ``belong in diplomacy, not bombs,'' capturing this pushback. In contrast, far fewer liberals shifted downward in the Decrease or Neutral conditions: 16 of 46 reduced their preference after interacting with the Decrease ChatBot, and only 9 of 68 did so after engaging with the Neutral ChatBot. However, this lower rate of observed change is partly attributable to a floor effect—many liberal participants had already selected the lowest value on the Likert scale, limiting the degree to which further downward movement was possible.

\subsection{Summary of Findings}

Our results point to a targeted impact of ChatBot driven persuasion on defense spending. Quantitatively, the one-way ANOVA on participants’ \textit{DefenseChange} scores (\textit{F}(2,335) = 8.93, \textit{p} $<$ 0.001) showed that both the \textbf{Decrease} and \textbf{Increase} ChatBots led to statistically significant shifts in the directions they promoted compared to the Neutral ChatBot. A Wilcoxon signed-rank test further indicated that within-group changes reached significance only for the \textbf{Decrease} group after Bonferroni correction, reinforcing that participants exposed to the Decrease ChatBot reduced their defense spending preferences more reliably than those in other conditions.

\paragraph{Group level persuasion patterns.} With the Decrease ChatBot, 16/46 liberals, 8/18 moderates, and 8/35 conservatives lowered their defense-spending preference. The Increase ChatBot raised support in 16/33 conservatives, 8/20 moderates, and 23/62 liberals, while most of the other liberals (32/62) backfired in their support for lower spending. Qualitative feedback showed liberals favored the ``reallocate-to-social-programs'' framing, conservatives aligned with fiscal-security rhetoric, and moderates shifted when efficiency gains were explicit. Thus, the Decrease ChatBot most reliably moved moderates, and value misalignment generally provoked resistance rather than persuasion.

\paragraph{Backfire Effect patterns. } Qualitatively, most participants shifted with the ChatBot’s framing, yet a notable subset backfired when faced with opposing persuasion. Among conservatives, 12/35 in the Decrease condition actually raised their defense budgets, often citing immigration or foreign conflicts. Likewise, 32/62 liberals resisted the Increase ChatBot, restating their anti-military stance and prioritizing education or healthcare. Moderates showed a softer version of this effect, usually holding steady or slightly nudging their original preference. Our Backfire effect aligns with Chan et al.’s observation that once people establish a position on a topic (misinformed or not), they tend to entrench themselves in that position and later resist attitude change; while Wood et al.'s observation cited the rarity of traditional backfire effect under correction, we observed greater prevalence of backfire effects under value-misalignment \cite{chanDebunk2017, Wood2019TheEB}.
\section{Discussion \& Future Work}
\label{sec:discussion}

Our work provides new evidence that value-framed AI interventions can meaningfully shape user attitudes on contentious policy issues. These effects are visible across both quantitative findings (Table \ref{tab:defense_data}) and qualitative responses.

\textbf{Interpretation of Findings.} One of the most consistent findings was that participants with moderate or slightly conservative leanings showed the greatest susceptibility to ChatBot framing. For instance, moderates in the Decrease condition were more likely to revise their opinions downward, especially when the ChatBot emphasized budgetary reallocation to domestic priorities. Similarly, slightly conservative participants exposed to the Increase ChatBot showed measurable shifts upward when the message focused on economic stability and national strength.

These interaction effects suggest that individuals with less rigid ideological commitments (or those whose values overlap with multiple political identities) may be more responsive to persuasive framing. In contrast, participants with strongly held views (particularly at ideological extremes) were more likely to resist persuasion or demonstrate a ``backfire'' effect, where the ChatBot's argument inadvertently reinforced, rather than shifted, their original stance. This aligns with prior research on value affirmation and motivated reasoning, and it highlights how persuasive AI can act not only as a shaping force, but also as a catalyst for entrenchment \cite{jakesch2023co, furumai2024, johnson2022ghost}. In contrast to previously observed correction-based backfire effects, our \emph{value-misalignment backfire effect} appears to be more prevalent and lends to more opportunities for LLM-based ChatBots to persuade and manipulate in general conversations about policy.

Our findings also point to important differences in how political groups respond to value-aligned persuasion. For instance, conservative participants interacting with the Decrease ChatBot were more likely to derail the conversation or reject the framing outright---often pivoting to issues such as immigration or government distrust. Liberal participants, in contrast, were more likely to resist the Increase ChatBot's push toward higher spending, often invoking social welfare or antiwar stances.
Moderates, however, frequently engaged with both perspectives and sometimes shifted their stance, suggesting that they may be especially vulnerable to subtle nudges. Our findings may provide guidance for each group when conducting future work in AI-mediated deliberation, particularly around de-escalation, consensus-building, or even polarization mitigation.

\textbf{Ethical Implications.} Our results suggest that value-framed ChatBots can steer people without announcing themselves as persuasive systems and that opacity matters. Participants generally trusted the assistant and often engaged it as a neutral aide wherein creating conditions where value cues can shape reasoning while remaining largely invisible \cite{Glickman2024HowHF}. This raises two related concerns: (1) \emph{informational asymmetry}, where the model knows it is framing while the user does not, and (2) \emph{misattribution of reasoning}, where users internalize arguments introduced by the model as if they were their own clarified views, rather than recognizing them as the model’s nudge. Together, these dynamics make it too easy to launder contested value tradeoffs through ostensibly objective conversation. Fisher et al. extend this point by showing that prior AI literacy can modestly blunt susceptibility, even when bias recognition itself does not, showing the role of user education in mitigation.

We therefore view our intervention as sitting squarely in a gray zone between decision support and attitudinal targeting. Unlike Fisher et al., who minimized prior attitudes by choosing obscure policy topics, our focus on defense spending engages a value-laden policy debate, raising different ethical stakes for hidden persuasion. Even though our prompts were simple and our tasks short, the measured shifts (and documented backfire among value-misaligned users) indicate that persuasive mechanisms do not require overt calls to action to be ethically consequential. Previous work similarly finds that partisan LLMs can alter opinions and allocations in the direction of the model's bias (even when the bias is contrary to the priors of the participants) while bias \emph{recognition} alone does little to blunt the effect \cite{fisher-etal-2025-biased, Carrella2025WarningPT}. Our data mirrors that pattern: trust plus plausibly helpful framing can be enough to move people. This combination of subtlety and efficacy is exactly what makes value-based steering hard to govern.

We also used deception where participants were not told the assistant would argue from a particular value-frame. Although our protocol was IRB-approved as a benign attitude intervention, the choice to conceal the persuasive intent introduces residual risks: transient attitude change, potential negative affect when bias is revealed, and a broader normalization of ``hidden persuasion'' in everyday tools. Future iterations should stress-test alternatives: soft disclosure that a range of normative perspectives will be presented, interface-level \emph{frame markers} that label value claims (e.g., \emph{Security frame}, \emph{Equity frame}) at the point of use, and user controls that allow opting into balanced or adversarially paired frames rather than a single persuasive arc. Informing the participants afterward by debriefing can also be a relevant strategy.

A second ethical tension concerns \emph{differential vulnerability}. Our quantitative and qualitative analyses point to groups who were more ``frame permeable,'' and to predictable failure modes (e.g., reactive entrenchment/backfire among strongly identified conservatives when speaking with the Decrease/Neutral ChatBot). Designing value-aligned assistants without acknowledging these asymmetries risks exacerbating polarization where the same nudge that helps one user articulate tradeoffs may harden another's threat schema. That suggests guardrails need to be audience-aware–not to microtarget harder, but to throttle persuasive intensity, diversify frames, or surface explicit counterarguments when signals of entrenchment appear. Our data suggest asymmetric susceptibility by group, but Fisher et al. document a clearer ceiling effect among aligned partisans, highlighting that vulnerability is not symmetric across all subgroups.

Finally, there is a platform level responsibility. Persuasive capacity scales faster than institutional oversight \cite{Glickman2024HowHF}. If models can steer preferences with light-touch framing in controlled studies, they can be repurposed for coordinated influence campaigns in the wild. That risk justifies commitments beyond bias ``filters,'' such as provenance for persuasive prompts (who authored the frame), auditable logs of value cues provided during the session, user-facing summaries at session end (\emph{an example being ``Frames emphasized today: Security, Efficiency, Patriotism''}), and friction for high stakes domains (politics, public health), where systems default to balanced multi-frame presentations or require explicit consent to enter a persuasive mode. Our study does not solve these governance questions; it makes their urgency harder to ignore.

In short, the ethical stakes are not limited to \emph{what} the assistant says, but \emph{how} and \emph{when} it says it, to whom and under what disclosures. We therefore advocate a research and design agenda that treats value transparency, user agency, and frame diversity as first-class features (not afterthoughts to accuracy) and that evaluates ``success'' not only by effect sizes but by whether people leave the interaction with a clearer view of the value tradeoffs they endorsed.

\textbf{Limitations and Future Work.} Our research focused on short-term interactions and did not assess how repeated or sustained exposure to value-driven ChatBots might affect long-term attitudes or trust in AI systems. Furthermore, our study was limited to one policy domain (defense spending) and one focal value (fiscal conservatism). Future research should expand this approach to other contested issues to assess generalizability of our findings. We inferred value alignment based on political self-identification and topic responses, but future research could incorporate validated scales for moral foundations, ideological rigidity, or value salience to better understand how belief structures interact with persuasive AI. Our sample, while diverse in political orientation, was drawn from participants already familiar with AI systems. Broader sampling is needed to capture the responses of less technologically literate populations, where trust cues can differ significantly. Finally, future work should explore value-aligned AI in more benevolent and collaborative contexts (such as public health, education, or civic engagement) to understand how value alignment can support end users.
\section{Conclusion}
\label{sec:conclusion}

We present a crowdsourced study with 336 participants interacting with LLM-driven ChatBots prompted to persuade by political stance (i.e., values-framing). Our quantitative analyses show that value-framed ChatBots can significantly influence participant attitudes. Our qualitative analysis of participants' conversations suggest that participants who hold moderate viewpoints and are open to the arguments of others seem to be more convinced by the Chatbots. Further, we find that people holding more extreme views may become entrenched in their rationale when ChatBots disagree with them, consistent with similar observations in psychology and misinformation research. Our exploratory findings identify some boundary conditions for backfire in value-framed ChatBot conversations and lay out testable predictions for replication and extension. Together, our results highlight potential ramifications of AI influence on civic discourse and attitude shifting, which can inform AI model and ChatBot designs in future research.

%\bibliography{reference}

\clearpage

\appendix

\begin{table*}[ht]
\centering
\caption{One-way ANOVA results for change in defense spending preference}
\label{tab:anova_defensechange}
\begin{tabular}{lcccc}
\hline
Source & Sum of Squares & df & F & $p$-value \\
\hline
Between Groups & 22.543 & 2 & 8.93 & $<0.001$ \\
Within Groups & 422.830 & 335 &  &  \\
\hline
\end{tabular}
\end{table*}

\begin{table*}[ht!]
\centering
\caption{Descriptive statistics for change in defense spending preference by ChatBot group. Highlight indicates significance.}
\label{tab:desc_defensechange}
\begin{tabular}{lccc}
\hline
\textbf{ChatBot Group} & \textbf{Mean} & \textbf{Std Dev} & \textbf{Count} \\
\hline
\rowcolor{yellow!25} Decrease & $-0.444$ & 1.136 & 99 \\
Neutral & $-0.252$ & 1.135 & 123 \\
Increase & $0.181$ & 1.100 & 116 \\
\hline
\end{tabular}
\end{table*}

\begin{table*}[ht]
\centering
\caption{Tukey's HSD post-hoc test for change in defense spending preference. Highlights indicate significance.}
\label{tab:tukey_defensechange}
\begin{tabular}{lccccc}
\hline
Group 1 & Group 2 & Mean Difference & $p$-adj & Lower & Upper \\
\hline
\rowcolor{yellow!25} Decrease & Increase & $-0.6255$ & $0.0002$ & $-0.9874$ & $-0.2636$ \\
Decrease & Neutral & $-0.1924$ & $0.4141$ & $-0.5495$ & $0.1647$ \\
\rowcolor{yellow!25} Increase & Neutral & $0.4331$ & $0.0087$ & $0.0908$ & $0.7754$ \\
\hline
\end{tabular}
\end{table*}

\begin{table*}[htbp]
\centering
\caption{Tukey's HSD post-hoc test for change in spending on defense. Highlights indicate significance.}
\label{tab:tukey_spendingdefense}
\begin{tabular}{lccccc}
\hline
Group 1 & Group 2 & Mean Difference & $p$-adj & Lower & Upper \\
\hline
\rowcolor{yellow!25} Decrease & Increase & $-0.6227$ & $0.0004$ & $-1.0035$ & $-0.2419$ \\
Decrease & Neutral & $-0.1638$ & $0.5607$ & $-0.5396$ & $0.2119$ \\
\rowcolor{yellow!25} Increase & Neutral & $0.4589$ & $0.0082$ & $0.0987$ & $0.8190$ \\
\hline
\end{tabular}
\end{table*}

\begin{table*}[htbp]
\centering
\caption{Wilcoxon signed-rank tests for defense spending preferences. Highlight indicates significance.}
\label{tab:wilcoxon_defense}
\begin{tabular}{lccc}
\hline
\textbf{Group} & \textbf{Statistic} & \textbf{$p$-value} & \textbf{Interpretation}\\
\hline
Overall & 3737.000 & 0.006 & Not Significant (after correction) \\
Increase & 413.500 & 0.100 & Not Significant \\
\rowcolor{yellow!25} Decrease & 130.000 & $<0.001$ & Significant \\
Neutral & 443.000 & 0.007 & Not Significant (after correction) \\
\hline
\end{tabular}
\end{table*}

\begin{table*}[htbp]
\centering
\caption{One-way ANOVA results for changes in other spending preferences. Highlight indicates significance.}
\label{tab:anova_otherspending}
\begin{tabular}{lcccc}
\hline
\textbf{Variable} & \textbf{Sum Sq} & \textbf{F} & \textbf{$p$-value} & \textbf{Significant} \\
\hline
\rowcolor{yellow!25} Spending on Defense & 23.029 & 8.24 & $<0.001$ & Yes \\
Spending on National Security & 4.082 & 1.681 & 0.188 & No \\
Spending on Crime Prevention & 5.709 & 2.59 & 0.077 & No \\
Spending on Public Schools & 0.396 & 0.293 & 0.746 & No \\
Spending on Science and Technology & 2.087 & 1.091 & 0.337 & No \\
Spending on Environmental Protection & 1.814 & 1.101 & 0.334 & No \\
\hline
\end{tabular}
\end{table*}

\end{document}